%% file: main.tex
\documentclass[runningheads]{llncs}
\usepackage{graphicx}
\input{packages}
\input{macros}

\input{abbreviations}
\begin{document}
\title{Transmitter Classification With Supervised Deep Learning}
%
%
\author{Cyrille~Morin\inst{1}\orcidID{0000-0001-5878-3501} \and
Leonardo~S.~Cardoso\inst{2,3}\orcidID{0000-0002-6647-6031} \and
Jakob~Hoydis\inst{2,3}\orcidID{0000-0002-0438-967X} \and
Jean-Marie~Gorce\inst{2,3}\orcidID{0000-0002-5389-0102} \and
Thibaud~Vial\inst{3}}
\authorrunning{C. Morin et al.}
%
\institute{Univ Lyon, Inria, INSA Lyon, CITI, France \\
\email{<first name>.<last name>@inria.fr}\and
Nokia Bell Labs, France \\
\email{jakob.hoydis@nokia-bell-labs.com}\\
\and
RTone, France\\
\email{thibaud.vial@rtone.fr}}
\maketitle              
\input{abstract}

\glsresetall

\input{introduction}

\section{The Identification Problem}\label{sec_identification_problem}

Current authentication schemes for packet transmission are based on transmitting an identification number inside of a frame header.
This poses two problems:

\begin{itemize}
	\item The number needs spectral and energy resources to be transmitted. 
	In \gls{iot} protocols, these resources are already very limited.
	\item The identification is trivial to spoof. 
	Meaning that the transmission needs to be authenticated again at higher levels with cryptographic means, or in cases where computing power or energy is limited it cannot be done.
\end{itemize}

\subsection{Characteristic elements of a point-to-point transmission}

\subsubsection{Channel effects}
First and most obvious is the impact of the channel between emitter and receiver.
In the standard case of a transmission with a fixed average power, the reception power is a direct indicator of the path loss and thus of the distance between emission and reception. 
This power can then serve as a coarse indicator of the emitter.
In this case, multi-path parameters can also be easily measured. These are highly dependent on the position of the emitter. 

These two elements have a high impact on the signal and are simple to measure, but they are dependent on the system topology which is expected to change over time, not on intrinsic properties of the radio cards. This could still be used as a way to detect impersonation by looking at sudden changes in these parameters, as studied in \cite{Xiao2009,Xiao2009a}.

\subsubsection{Power amplifier imperfections}
Homodyne radios suffer from IQ imbalance. 
The In-phase and Quadrature components of the signal do not go through the same path with the same components. 
This means that they may not be amplified by the same gain and the resulting constellation gets skewed. 
The second part of the imbalance comes from the fact that the two parts are not mixed with sine waves at exactly 90 degrees and the resulting constellation gets rotated.
This effect is used greatly in \cite{Sankhe2018} and \cite{Wong2018} but it's mostly present in devices with a homodyne architecture as those used in \gls{sdr} and not in superheterodyne radios that make up the majority of consumer devices.

Amplifiers of RF signal are non linear components. 
They are setup so they function mostly in their linear region but, even there, their response curve is not perfectly linear.
The parameters corresponding to this are slightly different from one radio chip to another, even amongst the same product line~\cite{Hanna2018}. 

\subsubsection{Local oscillator imperfections}
The local oscillator is tasked with translating the baseband signal to the carrier frequency.
It is however not able to set itself to the exact same frequency as the receiver.
This creates a frequency offset that may be characteristic of an emitter
The offset is not only dependent on intrinsic elements of the radio card, but also on temperature so it can change over time at a rate that can also be indicative of the emitter, if it's high enough to be measured.

The frequency translation operation that uses the oscillator to bring baseband signals to carrier frequency can present a local oscillator leak in homodyne devices.
In this case, a peak is sent at the oscillator frequency, whose power does not depend on the actual signal sent, but only on the amplifier settings.
The phenomenon is used in \cite{Hanna2018} to identify emitters even when, considering data sent, $ {SNR} = -\infty$.

\subsection{Avoiding dataset bias}

The aforementioned elements are not all desirable: The channel effects are only distinctive of the position of the emitter and may not be reliable when the environment becomes realistically dynamic. 
Some of the other imperfections are specific to homodyne radio cards. This means that they can appear in the USRPs that will be used, but not on many mainstream consumer grade superheterodyne radios. 
So reducing the reliance on these imperfections allows better generalisation of the results to other radio hardware architectures.

The USRPs are calibrated to remove IQ imbalance and DC offsets and the emission gain is set to reduce local oscillator leakage to a minimum. 
But channel effects cannot be removed as simply while still maintaining realistic over-the-air radio propagation (no cable). 
Instead of removing them, they can be randomised: 
When a parameter is random and varies over a range that overlaps these of the other emitters, one specific realisation of that random parameter cannot give insight on who is the transmitter.
In this case, the \gls{NN} will not learn to depend on that parameter to perform identification.

\input{cortexlab.tex}

\section{Dataset generation process}\label{sec_dataset_generation}

The data generation process is modular to allow for testing of a wide range of scenarios with various amount of bias.
There is a standard core process that can be parameterised to create different scenarios.
The signal processing chain is written as \gls{grc} flowgraphs for emission and reception that provide a light API for configuration.
All the high-level management of the experiments is done in small python and bash scripts for easy configuration and reproduction of scenarios.

\subsection{Core process}
There are 21 emitters, one receiver, and one scheduler.
Each of these, except the scheduler, uses a National Instruments USRP N2932 \gls{sdr} working at \SI{5}{\mega\samples\per\second} and \SI{433}{\mega\hertz}.
All the emitters have to transmit packets to the receiver and use the same frequency band. 
The scheduler's role is to ensure that there is no interference between packets from different emitters without needing to implement carrier sensing algorithm.
It also causes packets from different emitters to be sent in close temporal proximity.
This is useful if the environment is not static: one specific environment configuration could be indicative of a specific emitter if it was not the case.
Every millisecond the scheduler selects randomly one among the possible emitters, then sends a packet to it via UDP.
\begin{figure}[h]
	\centering
	\includegraphics[width=1\columnwidth]{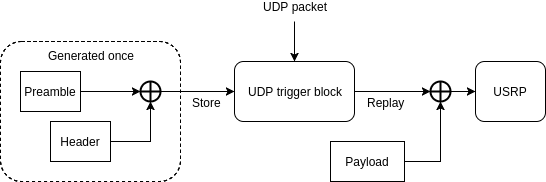}
	\label{fig:emitter_logic}
	\caption{Simplified emitter flowgraph}
\end{figure}

The emitters have their USRP set to burst mode. 
This means that when they are not actively transmitting, their amplifiers are off so they do not emit anything.
Even local oscillator leakage is prevented by this.
Upon reception of a scheduler packet, an emitter wakes up its USRP, waits for the amplifier to stabilise and sends a frame.
The frame is composed of three parts:
\begin{figure}[h]
	\centering
	\includegraphics[width=1\columnwidth]{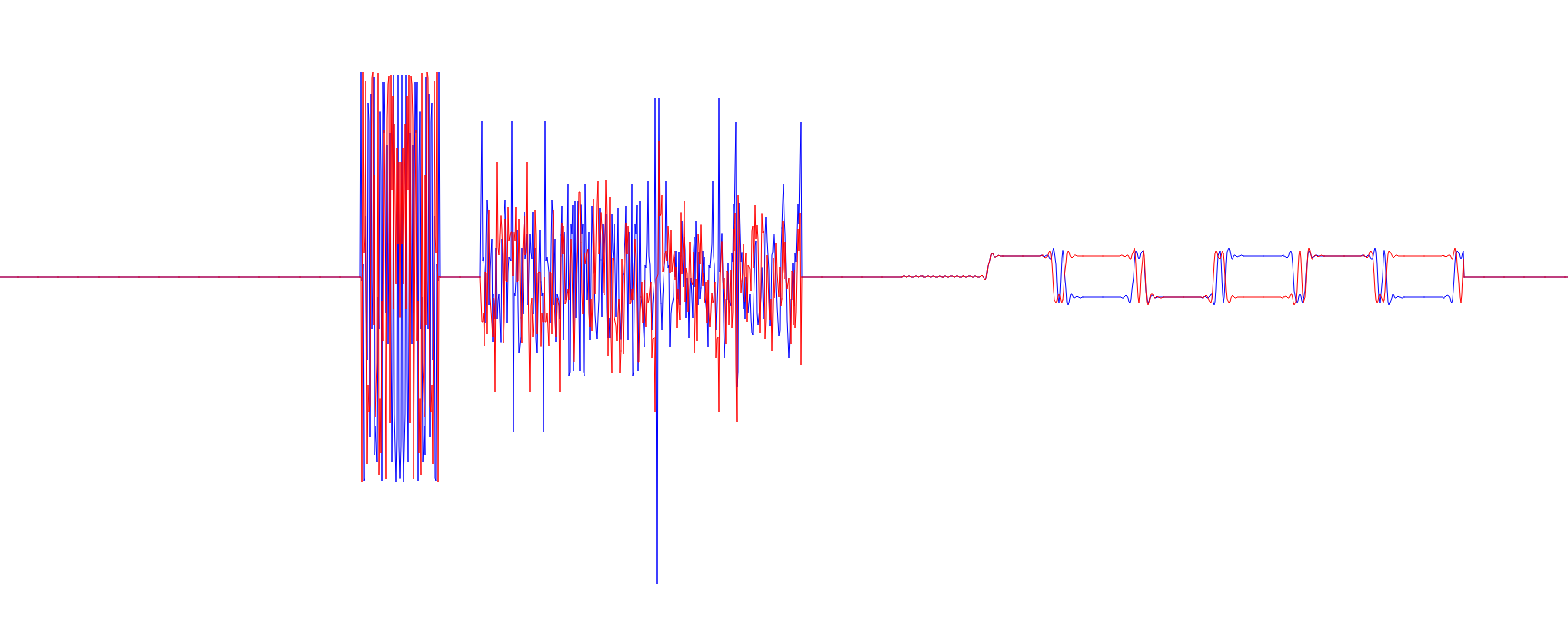}
	\label{fig:frame}
	\caption{Frame samples sent to USRP for emission, with zeroes for amplifier wake up, preamble, header, and payload with guard intervals}
\end{figure}

\begin{itemize}
	\item A known preamble for detection and time synchronisation with the frame.
	\item A header: an OFDM frame containing the identification of the emitter.
	\item The payload that we are interested in, containing either noise, a random or a static QPSK modulated sequence of 560 samples long.
	In all these cases, the payload does not contain any emitter specific information.
\end{itemize}
There is a time gap between the header and the payload to give the amplifier time to stabilise after the header and avoid interference between the two parts if there is a significant amplitude difference between the two parts.

\begin{figure*}[h]
	\centering
	\includegraphics[width=1\textwidth]{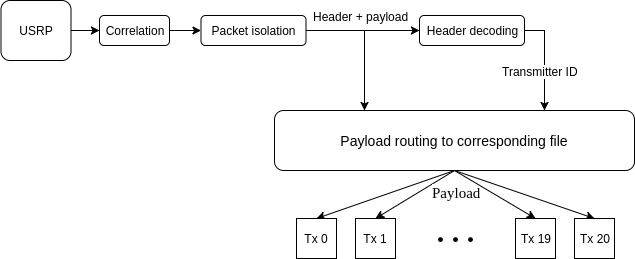}
	\label{fig:reception}
	\caption{Reception flowgraph}
\end{figure*}
The receiver's USRP stays on for the duration of the experiment.
It uses a correlator to detect the presence and the timing of the known preamble, then isolates the number of samples corresponding to the length of the header and the preamble.
An OFDM packet receiver is used to decode the header and the identification number is used to send the payload samples to be recorded in the corresponding file.

The grouping of the recorded files for one experiment in one scenario forms a dataset.
Experiment run times are set to gather about 50000 packets per emitter.

\subsection{Scenarios}

An experiment scenario has two main parameters to allow testing and selection of different elements:
\subsubsection{Type of signal}
The goal here is to identify the emitter based on how it sends data, not what it sends.
But every type of data does not necessarily yield the same classification accuracy.
To show this, three types of payload are tested:
\begin{itemize}
	\item A fixed sequence of QPSK modulated bits:
	All the emitters always send the exact same sequence: the bit sequence of the 802.15.4 preamble.
	This reduces the noise the \gls{CNN} has to deal with while still being a realistic case: here it's not the user data that is used but the preamble and this has to be transmitted anyways.
	\item A random sequence of modulated bits:
	The emitters generate random sequences of bits and these are modulated in the same way for every emitter. 
	QPSK is used as it is a commonly used modulation scheme for \gls{iot} devices.
	\item A noise sequence:
	The payloads are randomly uniformly generated from a noise source.
	This allows to test if the modulation choice has an impact on the performance or if any modulation would work.
	
\end{itemize}

\subsubsection{Transmission complexity}
\begin{itemize}
	\item Plain:
	The simplest of the modes.
	All the payloads from all emitters are sent with the same amplitude and nothing moves inside of the experimentation room.
	This is mostly used as a benchmark to compare the other channel randomising scenarios.
	\item Varying amplitude:
	The emission amplitude varies from one payload to another to emulate changes in path loss for each emitter without physically moving them. 
	This is implemented by scaling the IQ samples before sending them to the USRP, and not by changing the gain settings of the device because the amplifier takes a relatively long time to stabilise whereas scaling samples in software is instantaneous.
	Nothing moves inside of the room, so the multipath parameters are still static.
	\item Robot:
	The payloads are sent the same way as in the previous scenario.
	\begin{figure}[h]
		\centering
		\includegraphics[width=1\columnwidth]{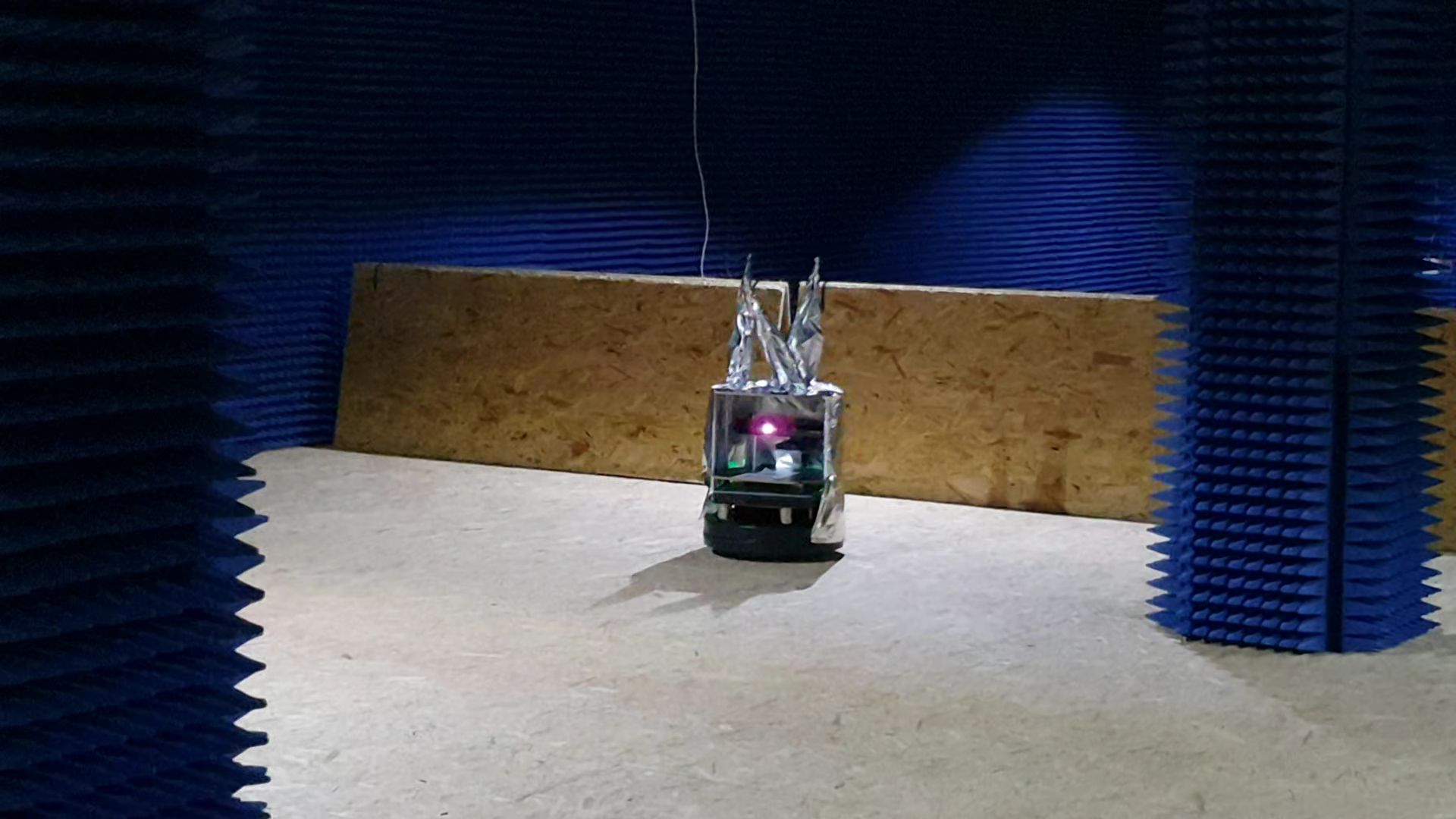}
		\label{fig:turtle}
		\caption{The Turtlebot robot with the metallic sheets, inside of \gls{cor} }
	\end{figure}
	A robot is introduced, covered with metallic sheets to increase radio waves reflections and set to move randomly inside the room.
	This introduces new and constantly changing reflections to randomise the multipath parameters.
\end{itemize}

\section{Learn to classify}\label{sec_classification}

\subsection{System architecture}
We use a \gls{CNN} type network with five layers of convolution and six dense layers. 
It takes 600 complex samples organised in a matrix of 600x2 float numbers for the Cartesian coordinates of the complex numbers.
And it outputs a vector of 21 numbers corresponding to the likelihood that the input was from one of the 21 transmitters.
\begin{figure}[h]
	\centering
	\includegraphics[width=1\columnwidth]{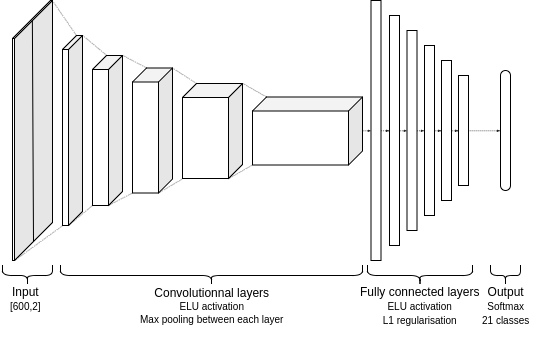}
	\label{fig:archi}
	\caption{Neural network architecture}
\end{figure}
Each layer has an \gls{elu} activation function except for the output layer which uses a $\softmax$ activation.

\subsection{Training phase}

Before training, datasets are randomly shuffled and then split into 3 parts: 70\% used for training, 10\% for validation and hyperparameter tuning and the last 20\% for testing.
Networks are trained over all the gathered datasets with the same architecture.
Training is done on mini-batches of 128 examples over more than 30 epochs, with an Adam optimiser, 0.001 of learning rate, l1 regularisation on the dense layers while minimising the categorical crossentropy loss function. 

Hyperparameter tuning was done by training networks with increasing amount of dense and convolutional layers, with varying batch sizes, learning rates and regularisation on a dataset that was found to be hard to train on: Varying amplitude and a payload of random bits.
They were trained for ten epochs and the best performing was selected.

\section{Results}\label{sec_results}

The first step is to establish a benchmark with the simplest scenario. 
This actually shows a comparison with the network in \cite{Sankhe2018} without artificial impairments. 
In \cite{Sankhe2018}, they use a very similar setup of 16 static USRPs placed in a room with no moving object.
Their network achieves a classification accuracy of \SI{98.6}{\percent} whereas the one studied here achieves \SI{99.9}{\percent} with more classes (21 instead of 16)

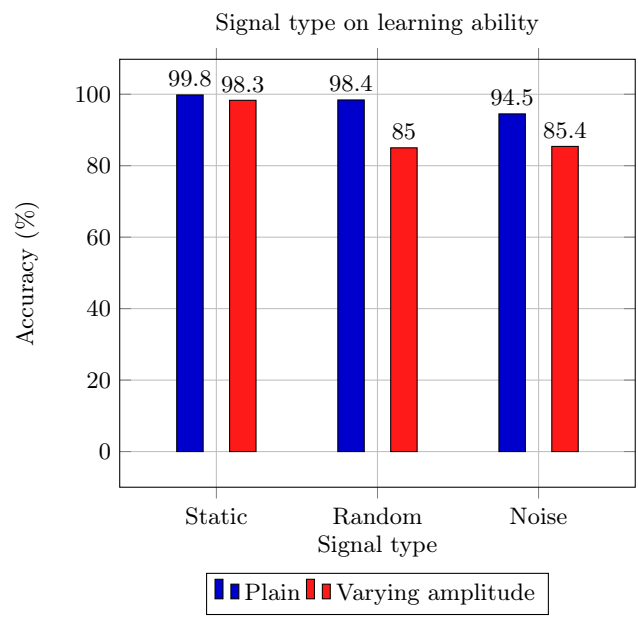
\begin{figure}[]
	\centering
	\begin{tikzpicture}
	\begin{axis}[ybar=10pt,
	title=Signal type on learning ability, 
	xlabel={Signal type},ylabel={Accuracy (\%)}, 
	xtick=data, symbolic x coords={Static, Random, Noise, Rab}, 
	legend style={at={(0.5,-0.2)},anchor=north,legend columns=-1},
	enlarge x limits=0.3,
	enlarge y limits=0.1,
	grid=major,
	nodes near coords, nodes near coords align={vertical},
	ymin=0,]
	\addplot [fill=blue!80!black] table[
	y={Plain},]
	{graph_data_type.txt};
	\addplot [fill=red!90] table[
	y={Var},]
	{graph_data_type.txt};
	
	\legend{Plain, Varying amplitude}
	\end{axis}
	\end{tikzpicture}
	\caption{Accuracy reached by networks trained on plain or varying amplitude scenarios and with various signal types. The accuracy is measured on the test set from the training dataset.}
	\label{fig:Signal_type}
\end{figure}

Fig.~\ref{fig:Signal_type} presents the classification accuracy achieved for the three signal types, while comparing the Plain and Varying amplitude scenarios.
Having randomness in the transmitted signal degrades classification accuracy and this degradation increases with the scenario complexity.
However, we can also see that having a completely random noise does not cause a significant performance loss over a QPSK modulated signal, even though the latter has a limited constellation size compared to noise.
Thus a higher order modulation should not cause further accuracy losses, but the ideal case is to use a static signal, as would be a frame preamble.

\begin{figure}[]
	\centering
	\begin{tikzpicture}
	\begin{axis}[ybar=10pt,
	title=Dependency on channel state, 
	xlabel={Training scenario},ylabel={Accuracy (\%)}, 
	xtick=data, 
	symbolic x coords={Plain, Varying, Robot, Rab}, 
	enlarge x limits=0.3,
	enlarge y limits=0.1,
	nodes near coords, nodes near coords align={vertical},
	legend style={at={(0.5,-0.2)},anchor=north,legend columns=-1},
	grid = major,
	ymin=0,]
	\addplot [fill=blue!80!black] table[
	y={Train},]
	{graph_training_ability.txt};
	\addplot [fill=red!90] table[
	y={Chair},]
	{graph_training_ability.txt};
	
	\legend{Training, Resilience to channel change}
	\end{axis}
	\end{tikzpicture}
	\caption{Accuracy of networks trained one one scenario with static signal and tested, either on test data from the training dataset or on a dataset with the same scenario but with a chair added to the room.}
	\label{fig:Train_test}
\end{figure}
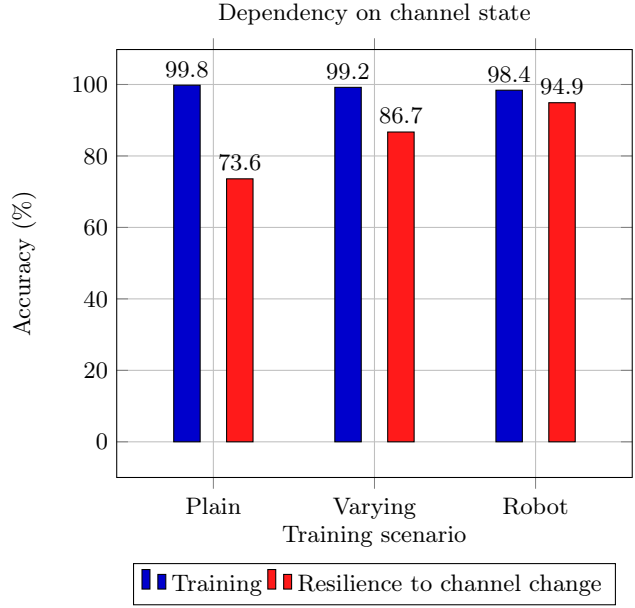

Fig.~\ref{fig:Train_test} studies the impact of environment modification on classification accuracy for the three scenarios. 
In it, one can observe that, the more complex a scenario is, the harder it is for a network to train on it, by a slight margin but the better it is at resisting changes in the environment. 
The training datasets were generated one after another, with no changes inside the \gls{cor} room, then a metallic stool was introduced inside the room and the other datasets were generated.
This ensures that the change in signal propagation is exactly the same for the 3 scenarios.
The Plain scenario suffers from a big loss of accuracy, as was noticed in \cite{Sankhe2018}, but the Robot one is very resilient to this kind of change.

\begin{figure}[]
	\centering
	\begin{tikzpicture}
	\begin{axis}[ybar=9pt,
	title=Generalisation to other scenarios, xlabel={Training scenario},ylabel={Accuracy (\%)}, xtick=data, symbolic x coords={Plain, Varying, Robot, Rab},
	enlarge x limits=0.3,
	enlarge y limits=0.1,
	nodes near coords, nodes near coords align={vertical},
	legend style={at={(0.5,-0.2)},anchor=north,legend columns=-1},
	ymin=0,
	grid=major,
	xminorgrids=true,
	xmajorgrids=false,
	minor x tick num=1,]
	\addplot [fill=blue!80!black] table[
	y={Plain},]
	{graph_cross_test.txt};
	\addplot [fill=red!90] table[
	y={Var},]
	{graph_cross_test.txt};
	\addplot [fill=yellow!90] table[
	y={Robot},]
	{graph_cross_test.txt};
	
	\legend{Plain, Varying amplitude, Robot}
	\end{axis}
	\end{tikzpicture}
	\caption{Accuracy of networks trained on one scenario and tested on the others. Here the signal is of the static type and the environment is untouched.}
	\label{fig:Cross_test}
\end{figure}
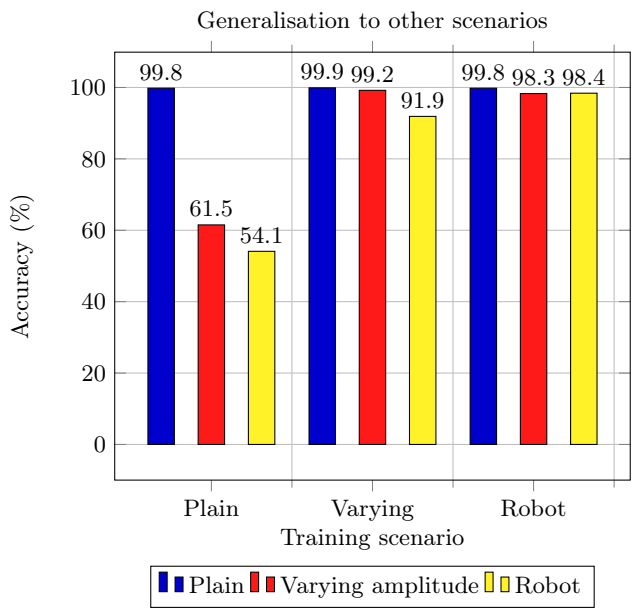

Finally, Fig.\ref{fig:Cross_test} focuses on the ability to generalise to other scenario types.
A network is trained on each scenario and tested on all of them.
We can observe a decrease in accuracy when a simple scenario is tested on a more complex one.
From this, one can infer that the more random the channel is, the more the trained network is able to cope with a change of scenario and also with a change in the environment.

The key takeaway from these results is as follows: 
if one were to implement a transmitter identification system in a production setting, its goal should be to train the neural network with the maximum amount of channel variability.

\input{conclusion}

\section*{Acknowledgement}
This work was supported by Inria Nokia Bell Labs ADR “Analytics and machine learning for mobile networks".
Experiments presented in this paper were carried out using the FIT/CorteXlab testbed. (see \url{http://www.cortexlab.fr}).

%
%
%
 \bibliographystyle{splncs04}
 \bibliography{bib_abrv,bibliography}

\end{document}

%% file: packages.tex
\usepackage[T1]{fontenc}
\usepackage[utf8]{inputenc}
\usepackage{mathtools}
\usepackage{algorithm}
\usepackage[x11names]{xcolor} 
\usepackage{tikz}
\usepackage{pgfplots}
\usepackage{glossaries}
\usepackage{siunitx}
\usepackage{hyperref}

%% file: macros.tex
\newcommand{\softmax}{\mathop{\mathrm{softmax}}}

\DeclareSIUnit{\samples}{sample}


%% file: abbreviations.tex
\newacronym{ACM}{ACM}{adaptive coding and modulation}
\newacronym{ADC}{ADC}{analog-to-digital conversion}
\newacronym{AGC}{AGC}{automatic gain control}
\newacronym{AWGN}{AWGN}{additive white Gaussian noise}
\newacronym{BER}{BER}{bit error rate}
\newacronym{BLER}{BLER}{block error rate}
\newacronym{BP}{BP}{backpropagation}
\newacronym{BPTT}{BPTT}{backpropagation through time}
\newacronym{CE}{CE}{cross-entropy}
\newacronym{CFO}{CFO}{carrier frequency offset}
\newacronym{CNN}{CNN}{convolutional neural network}
\newacronym{cor}{FIT/CorteXlab}{Future Internet of Things / Cognitive Radio Testbed~\cite{massouri2014cortexlab}}
\newacronym{CSI}{CSI}{channel state information}
\newacronym{DAC}{DAC}{digital-to-analog conversion}
\newacronym{DL}{DL}{deep learning}
\newacronym{DFT}{DFT}{discrete Fourier transform}
\newacronym{elu}{ELU}{exponential linear unit}
\newacronym{FFT}{FFT}{fast Fourier transform}
\newacronym{GAN}{GAN}{generative adversarial network}
\newglossaryentry{grc}{name={GNU Radio}, description={}}
\newacronym{GRU}{GRU}{gated recurrent unit}
\newacronym{iid}{i.i.d.\@}{independent and identically distributed}
\newacronym{IFFT}{IFFT}{inverse fast Fourier transform}
\newacronym{iot}{IoT}{Internet of Things}
\newacronym{KL}{KL}{Kullback-Leibler}
\newacronym{LSTM}{LSTM}{long short-term memory}
\newacronym{MDP}{MDP}{Markov decision process}
\newacronym{ml}{ML}{machine learning}
\newacronym{MLP}{MLP}{multilayer perceptron}
\newacronym{MIMO}{MIMO}{multiple-input multiple-output}
\newacronym{MSE}{MSE}{mean squared error}
\newacronym{NN}{NN}{neural network}
\newacronym{OFDM}{OFDM}{orthogonal frequency-division multiplexing}
\newacronym{pdf}{pdf}{probability density function}
\newacronym{pmf}{pmf}{probability mass function}
\newacronym{PSNR}{PSNR}{Peak Signal to Noise Ratio}
\newacronym{RBF}{RBF}{Rayleigh block-fading}
\newacronym{ReLU}{ReLU}{rectified linear unit}
\newacronym{rf}{RF}{radio frequency}
\newacronym{RL}{RL}{reinforcement learning}
\newacronym{RNN}{RNN}{recurrent neural network}
\newacronym{sdl}{SDL}{supervised deep learning}
\newacronym{sdr}{SDR}{software refined radio}
\newacronym{SFO}{SFO}{sampling frequency offset}
\newacronym{SGD}{SGD}{stochastic gradient descent}
\newacronym{SINR}{SINR}{signal-to-interference-plus-noise ratio}
\newacronym{SNR}{SNR}{signal-to-noise ratio}
\newacronym{wrt}{w.r.t.\@}{with respect to}

%% file: abstract.tex
\begin{abstract}
Hardware imperfections in RF transmitters introduce features that can be used to identify a specific transmitter amongst others.
Supervised deep learning has shown good performance in this task but using datasets not applicable to real world situations where topologies evolve over time. 
To remedy this, the work rests on a series of datasets gathered in the \gls{cor} to train a \gls{CNN}, where focus has been given to reduce channel bias that has plagued previous works and constrained them to a constant environment or to simulations. 
The most challenging scenarios provide the trained neural network with resilience and show insight on the best signal type to use for identification, namely packet preamble.
The generated datasets are published on the Machine Learning For Communications Emerging Technologies Initiatives web site\footnote{\href{https://wiki.cortexlab.fr/doku.php?id=tx-id}{Datasets and usage and generation scripts can also be found there: https://wiki.cortexlab.fr/doku.php?id=tx-id}} in the hope that they serve as stepping stones for future progress in the area.
The community is also invited to reproduce the studied scenarios and results by generating new datasets in \gls{cor}.
\keywords{Transmitter Identification  \and RF fingerprinting \and Deep Learning.}
\end{abstract}

%% file: introduction.tex
\section{Introduction}

Communication systems' constant evolution requires a constant search for new techniques that allow to squeeze out every bit of performance out of the system, a constant need to improve spectral efficiency in order to achieve the theoretical maximum capacity given by Shannon's law. This requirement is all the more important in systems that transmit many small packets, like \gls{iot}, where currently headers may outweigh the number of payload bits transmitted. Furthermore, headers are currently the only barrier against transmitter identification errors and transmitter impersonation on edge devices that don't have the ressources to use cryptographic protocols. Indeed, security issues are of utmost importance in this new era where hackers are able to easily attack transmissions even at the physical layer. Hence, a new means to provide secure identification of transmissions is needed, both to improve security as well as to render headers a thing of the past.

In the last five years, \gls{sdl} has imposed itself as the tool to achieve state-of-the-art performance in many fields, starting with image processing to voice recognition, product suggestion, and more generally data analysis and signal processing in physics, medicine and consumer products.
\gls{sdl} really shines in cases where labelled data is plentiful and mathematical models are not known. 
In the radio communication world, data is generated by the Terabyte per second all over the world, but for traditional applications, precise models already exist.
The existence of those good models explains why \gls{sdl} is not yet widely used in radio communication.
Yet it's starting to gain traction in the last couple of years, especially with channel decoding~\cite{Nachmani2016} and spectrum monitoring~\cite{OShea2016}.
Indeed, the tool promises increased performance in areas where models are yet to be derived and algorithms are not yet practical for real time implementations.
The task of identifying transmitters has attracted some attention in the last years, with two different approaches:

First, the confirmation of identity, by comparing a received signal with a previously authenticated one to verify if those characteristics match. 
This approach facilitates handling of changing environment by constantly updating the stored authenticated signal, as long as transmissions are more frequent than channel variations.
The fingerprints for each device is not stored inside of the identification system but in the recorded signals.
It can allow the identification of emitters unseen at training time, but increases processing times: comparisons need to be made with every known device.
In \cite{Xiao2009a}, the authors study channel responses in a simulated building floor to emulate spatial variations. 
In \cite{Weinand2017}, channel response is also used, but this time using a Gaussian Mixture Model to study similarities between samples gathered on real radio devices.

In second come the classifiers, and these are the more recent and numerous works, where a system is tasked not to give a similarity score between current and previous samples but to directly output the identity of the transmitter amongst a pool of previously seen radios.
A \gls{CNN} is used in \cite{Wong2018} to estimate IQ imbalance parameters of incoming signal from a simulation environment and these parameters are used in a classical Bayesian decision process.
Then \cite{Chatterjee2018} leverages parameters estimated routinely by decoding systems: DC and frequency offsets, IQ imbalance and channel information as input to a neural network. Matlab simulations show 99\% accuracy with  up to \num{10000} devices.
In \cite{Merchant2018} the authors use 7 consumer Zigbee transmitters to gather data. 
They remove the decoded data from the received signal to isolate channel and transmitter information effects before feeding it to a \gls{CNN} and achieve up to 90\% classification accuracy. 
However, one could argue that some hardware effects depend on the transmitted signal, for example amplifier non linearities, so removing it could be detrimental to the identification process.
Different machine learning techniques are evaluated in \cite{Youssef2017} over a dataset of signals gathered with real USRP devices accounting for 6 radio interfaces. 
They develop an architecture called a multi layer perceptron made of a network of small neural networks and introduce the use of wavelet transforms with the goal of learning to classify with as few training samples as possible.
The work in \cite{Hanna2018} focuses on amplifier characteristics and measures non linearity variations between 7 USRP cards. The data is used to train a network on hundreds of simulated devices. 
It also shows the effects of local oscillator leakage on classification accuracy.
Finally, in \cite{Sankhe2018} attention is switched to IQ imbalance and DC offset to perform classification.
A \gls{CNN} is trained on dataset made with 16 different USRP devices but it is not able to cope with changes in environment between experiments.
For this reason, artificial impairments are added to the signal and increase classification accuracy.
Still in the same context, the work in~\cite{Avatefipour2018} deals with wired communications to increase security and prevent intrusion of malicious systems in the network inside a car.

As previously stated, some works base the identification of transmitters on characteristics outside of the scope of the transmitter radios themselves using, for example, the channels~\cite{Xiao2009a,Weinand2017}. In realistic applications, however, radios are rarely at fixed locations and channel characteristics evolve with the surrounding environment. A better identification method would be to base the identification on the \gls{rf} \emph{signature} of the radios themselves, as addressed in~\cite{Wong2018,Hanna2018,Chatterjee2018,Sankhe2018}. However, in those works, either simulations or simple datasets were used which may not provide a biased free classifier that actually focuses on \gls{rf} signatures. Furthermore, in those works, no attention is given on the quality of the dataset itself for the identification task at hand (absence of bias, reproducibility, interference with outside sources), and in~\cite{Sankhe2018}, artificial impairments remove possible security claims: if identification is done on software added elements, any malicious software can do the same and impersonate the user.

To counter the problems encountered by previous works, herein, an extensive dataset campaign was generated aiming to train a more robust classifier. The datasets were carefully crafted to avoid biases like channel, transmitted power, packet structure, and receiver position through different measurement campaigns in a controlled environment. 
We use the \gls{cor} to gather experimental datasets to train neural network classifiers able to identify emitters based on their hardware characteristics.
The datasets are available online, as well as the scripts used to generate them, so that anyone can reproduce them on \gls{cor}.

The remainder of this paper is organised as follows. Section~\ref{sec_identification_problem} deals with the identification problem itself and the characteristics that make it particular. It also describes the \gls{cor} testbed, used for the measurement campaign. Section~\ref{sec_dataset_generation} describes the dataset generation process required to train a signature based \gls{CNN}. Then, in Section~\ref{sec_classification} the \gls{CNN} structure is described as well as the training process. Results are presented and discussed in Section~\ref{sec_results}. Finally, conclusions and perspectives are drawn in Section~\ref{sec_conclusions}.

%% file: cortexlab.tex
\subsection{The \Gls{cor} platform for learning}

The \gls{cor} testbed is a \gls{sdr} testing facility in France that enables testing of physical (and higher) layer techniques for future wireless systems. It counts with 42-node high-performance \gls{sdr} nodes, whose frequency range is roughly from 400~MHz to 4~GHz, at 20~MHz of maximum bandwidth. It is fully accessible to the wireless research community and uses GNU Radio as its programming environment. One of its main characteristics, the one that makes \gls{cor} particularly well suited for \gls{ml}, is its shielded room. It allows for reproducible experiments since the shielding provides a fully a controlled environment. Also important for \gls{ml} is the presence of a server equipped with GPUs connected via high speed data links to the \gls{sdr} nodes, which allow on-the-fly training and exploitation of \gls{sdl} techniques.



%% file: conclusion.tex
\section{Conclusions}\label{sec_conclusions}

The task of identifying transmitters based on hardware physical characteristics and imperfections has been gathering attention in the literature recently.
These identification strategies have been based aspects such as IQ imbalance, amplifier non linearities or channel properties.
Most of the works in the area involve the use of machine learning, and more specifically deep learning with good identification accuracies.
However, the data used to train the neural networks from prior works can not guarantee bias avoidance against unwanted elements, such as channel effects.

In this work, instead of focusing on a specific radio imperfection, neural networks were trained on raw IQ samples so as not to overlook any effect.
On the other hand, data was generated with the goal of minimising the role of the unwanted channel on the identification task by randomising the various channel parameters.

The considered neural network architecture shows state-of-the-art performance when tested on similar scenarios as previous works.
An exploration of various parameters was done with results that shows that the identification task is simpler when transmitted signals do not change and that unknown signals only incur a small performance decrease, independently of the modulation type.
They also show that training data with increased complexity do not impede learning ability but provides increased robustness against environment modifications.
Finally, from the experiments and the results, we see that \gls{cor} provides the stability and reproducibility necessary for machine learning approaches.

We plan to extend this setup to use packets recorded from not just one but from many receivers to increase the randomness of the perceived channel with the transmitters and evaluate the robustness of this approach to environment modification.
Another direction of exploration is to verify that higher order modulation schemes do not impair classification accuracy with respect to what observed with QPSK.